\documentstyle[epsfig,12pt]{article}

\newcommand{\bce}{\begin{center}} 
\newcommand{\ece}{\end{center}}
\newcommand{\beq}{\begin{equation}}
\newcommand{\eeq}{\end{equation}}
\newcommand{\bea}{\vspace{0.25cm}\begin{eqnarray}}
\newcommand{\eea}{\end{eqnarray}}

\newcommand{\bk}{{\bf k}}

\newcommand{\br}{{\bf r}}

\newcommand{\bm}[1]{\mbox{\boldmath $#1$}}

\newcommand{\ba}{\begin{array}}
\newcommand{\ea}{\end{array}}

\newcommand{\doublespace}{
    \renewcommand{\baselinestretch}{1.6}\large\normalsize}

\newcommand{\bkappa}{\mbox{\boldmath ${\kappa}$}}
\newcommand{\bPhi}{\mbox{\boldmath ${\Phi}$}}

\def\lsim{\mathrel{\rlap{\lower4pt\hbox{\hskip1pt$\sim$}}
    \raise1pt\hbox{$<$}}}         
\def\gsim{\mathrel{\rlap{\lower4pt\hbox{\hskip1pt$\sim$}}
    \raise1pt\hbox{$>$}}}         

\def\Pom{{\bf I\!P}}

\def\beq{\begin{equation}}
\def\endeq{\end{equation}}
\def\arr{\begin{eqnarray}}
\def\endarr{\end{eqnarray}}

\makeindex

  \advance\oddsidemargin  by -1in
  \advance\evensidemargin by -1in
  \advance\topmargin      by -0.5in
\begin{document}

\vspace{2.0cm}

\begin{flushright}
\end{flushright}

\vspace{1.0cm}

\begin{center}
{\Large \bf 
Heavy quark currents  in Ultra-High Energy Neutrino 
Interactions} 

\vspace{1.0cm}

{\large\bf R.~Fiore$^{1 \dagger}$ and V.R.~Zoller$^{2 \ddagger}$}

\vspace{1.0cm}

$^1${\it Dipartimento di Fisica,
Universit\`a     della Calabria\\
and\\
 Istituto Nazionale
di Fisica Nucleare, Gruppo collegato di Cosenza,\\
I-87036 Rende, Cosenza, Italy}\\
$^2${\it
ITEP, Moscow 117218, Russia\\}
\vspace{1.0cm}
{ \bf Abstract }\\
\end{center}
We discuss heavy quark contributions to the neutrino-nucleon total cross 
section at very high energies,  well above the real 
 top  production threshold.
The top-bottom weak  current is found to  generate strong 
left-right asymmetry of neutrino-nucleon interactions.
 We separate contributions of 
different helicity states and  make use of the 
$\bkappa$-factorization  to derive simple and 
practically  useful formulas for the left-handed ($F_L$) and 
right-handed ($F_R$)
 components of the conventional structure function 
 $2xF_3=F_L-F_R$ in terms of the integrated  gluon density. 
We show that $F_L\gg F_R$ and, consequently,
$xF_3\approx F_T$, where $F_T$ is the transverse structure function. 
The conventional  structure function $F_2=F_S+F_T$ at 
$Q^2\ll m_t^2$ appears to be dominated by its  scalar 
(also known as longitudinal) component  $F_S$
and the hierarchy $F_S\gg F_L\gg F_R$ arises naturally.
 We  evaluate the total  neutrino-nucleon  cross section 
at ultra-high energies  within the color 
dipole BFKL formalism.

\doublespace

\vskip 0.5cm \vfill $\begin{array}{ll}
^{\dagger}\mbox{{\it email address:}} & \mbox{fiore@cs.infn.it} \\
^{\ddagger}\mbox{{\it email address:}} & \mbox{zoller@itep.ru} \\
\end{array}$

\pagebreak



The neutrino astrophysics \cite{NeuAstro} deals with 
neutrinos of ultra-high energy
(UHE), well above the threshold of the top quark production.
The role of heavy quark currents in  UHE neutrino
interactions has been explored extensively \cite{HQ}. 
 Different approaches
to including the top quark effects in the UHE neutrino-nucleon  
cross section $\sigma^{\nu N}$  based on \cite{HQ} are in use \cite{tb}. 
 The goal of this communication is
to show that a specific
choice of relevant  degrees of freedom greatly simplifies the analysis of
$\sigma^{\nu N}$ in terms
of the nucleon structure functions (SF). Making use of the 
$\bkappa$-factorization
we  isolate the leading contributions involving
 big log's of $Q^2$, $m_t^2$ and $1/x$ to  derive
 simple and  practically useful formulas for 
the UHE  
neutrino-nucleon SFs.

The UHE interactions correspond to neutrino  energy above 
$E_{\nu}\sim 10^6$ GeV. The overall hardness
 scale of the process 
\beq
\nu N\to \mu X
\label{eq:PROC}
\eeq
 is determined by the gauge boson mass
\beq
Q^2\sim m_W^2.
\label{eq:Q2W2}
\eeq
Therefore,   the W- boson exchange   probes the gluon 
density in the target  nucleon at 
very small values of  Bjorken $x$. Consequences  of this observation 
for $\sigma^{\nu N}(E_{\nu})$ are widely  discussed \cite{NeuSec}
(for most recent publications  see \cite{Thorne}).

At small $x$, to the Leading Log$({1/x})$ approximation, 
it is legitimate to consider the $W$-nucleon scattering in the 
laboratory frame
in terms of interactions with the target of the $q\bar q\prime$-pair 
(color dipole) which 
the light-cone W-boson transforms into at large upstream distances.
  The 
dynamics  of  the $\log(1/x)$-evolution \cite{BFKL} 
of  the   dipole-nucleon cross section is  described by the 
infrared regulated  BFKL equation with  running coupling 
 derived in \cite{NZZJL94,NZJETP94}.

The differential  cross section for the $\nu N$ interactions
 is expressible  in terms of the
scalar (also known as longitudinal), 
left-handed and right-handed
structure functions denoted by $F_S$, $F_L$ and  $F_R$,  respectively.
\beq
x{d\sigma^{\nu N}\over dxdQ^2}={G_F^2\over 4\pi}
\left({m_W^2\over m_W^2+Q^2} \right)^2
\left[2(1-y)F_S+ F_L + (1-y)^2F_R\right].
\label{eq:DSDXDY}
\eeq
The structure functions $F_{\lambda}$ are
 related to the absorption cross sections 
 for the $W$-boson in the polarization state $\lambda=S,L,R$,
\beq
F_{\lambda}(x,Q^2)={Q^2\over 4\pi^2\alpha_W}\sigma_{\lambda}(x,Q^2),
\label{eq:FTOSIG}
\eeq 
where $\alpha_W=g^2/4\pi$, $g^2=G_Fm_W^2/\sqrt{2}$
and the $W\to t\bar b$ transition vertex is
$$
g\bar t\gamma_{\mu}(g_V-g_A\gamma_5)b.
$$
In terms of $F_L,F_R$ the transverse structure function $F_T=F_2-F_S$
reads  $F_T=(F_L+F_R)/2$, and 
left-right antisymmetric structure function $F_3$ is as follows 
$2xF_3=F_L-F_R$. Switching $L$ to $R$ in Eq.(\ref{eq:DSDXDY})
 yields 
the $\bar\nu N$  cross section.

Then, making use of  the color dipole factorization \cite{NZDLLA} we 
arrive at  the 
leading-$\log(1/x)$
$\bkappa$-factorization 
representation for the   absorption cross section for 
the $W$-boson in the 
polarization state $\lambda$ 
\bea
\sigma_{\lambda}(x,Q^{2})=\int_0^1 dz \int d^{2}\br|\Psi_{\lambda}(z,r)|^2
\sigma(x,r)\nonumber\\
= {\alpha_{W}\over \pi}\int_0^1 dz \int d^{2}\bk
 \int { d^{2}\bkappa 
\over \bkappa ^{4}}\alpha_{S}(q^{2})
{\cal F}(x_g,\bkappa ^{2})\left(S_{\lambda}\Phi_0^2+P_{\lambda}\bPhi_1^2\right).
\label{eq:FACT}
\eea

We address first the second line of Eq.(\ref{eq:FACT})
{\footnote {To compare Eq.(\ref{eq:FACT}) with  Eq.(2) of Ref. \cite{BGNPZ} 
one should make the 
 substitution:  
$${4\alpha_S\over \pi}{V(\kappa)\kappa^4\over 
(\kappa^2+\mu_G^2)^2}\to {\cal F}(x_g,\bkappa^2).$$
The numerical factor $16$  in Eq.(2) of Ref. \cite{BGNPZ} 
should be  understood as $4$.}}
 to derive 
 some useful  analytical expressions  for 
$F_{\lambda}(x,Q^2)$. The first line of 
Eq.(\ref{eq:FACT}) will be used in our numerical studies of $
\sigma^{\nu N}(E_{\nu})$. 

No restrictions on the quark transverse momentum,  $\bk$,
and the gluon transverse momentum, $\bkappa$ are imposed 
in Eq.(\ref{eq:FACT}), where
$${\cal F}(x_g,\bkappa ^{2})=
{\partial G(x_g,{\bm \kappa}^2)\over \partial \log\kappa^2},$$
and  $G(x_g,{\bm \kappa}^2)=x_gg(x_g,{\bm \kappa}^2)$ is the 
integrated gluon structure function. 
It is taken at
\beq
x_g={{Q^2+M^2}\over {W^2+Q^2}}
\label{eq:XG}  
\eeq
and the transverse mass of the $t\bar b$ pair is
\beq
M^2={{m_t^2+{\bk}^2}\over {z}}+
{{m_b^2+({\bk}-{\bkappa})^2}\over {1-z}},
\label{eq:M2}  
\eeq
where  $z$ is the fraction of the
$W$-boson
light cone momentum carried by the $t$-quark.

The terms proportional to $S_{\lambda}$ and $P_{\lambda}$ 
describe  interaction of the 
quark-antiquark states with the angular momentum $L=0$ (S-wave) and $L=1$
(P-wave), respectively.
The S-wave  and the P-wave  factors  in 
Eq.(\ref{eq:FACT}) for the right-handed and  left-handed $W$-bosons 
 are as follows \cite{FZ1} 
\bea
S_R=\left(g_V\left[(1-z)m_t+zm_b\right]
+g_A\left[(1-z)m_t-zm_b\right]\right)^2\nonumber\\
P_R=(g_V-g_A)^2z^2+(g_V+g_A)^2(1-z)^2\nonumber\\
S_L=\left(g_V\left[(1-z)m_t+zm_b\right]
-g_A\left[(1-z)m_t-zm_b\right]\right)^2\nonumber\\
P_L=(g_V-g_A)^2(1-z)^2+(g_V+g_A)^2z^2
\label{eq:SPLR}
\eea
and for the scalar/longitudinal polarization \cite{Kolya92,FZ1}
\bea
S_S=(g_V^2/Q^{2})\left[2Q^2z(1-z)+(m_t-m_b)
\left[(1-z)m_t-zm_b\right]\right]^2\nonumber\\
+(g_A^2/Q^{2})\left[2Q^2z(1-z)+(m_t+m_b)
\left[(1-z)m_t+zm_b\right]\right]^2\nonumber\\
P_S=(g_V^2/Q^{2})(m_t-m_b)^2+(g_A^2/Q^{2})(m_t+m_b)^2
\label{eq:SPS}
\eea
In the  charged current  interactions  $g_A=g_V=1$. 
In the neutral current neutrino interactions $m_q=m_{\bar q^{\prime}}$ and 
corresponding vector and axial-vector couplings are given by 
the Standard Model.

In   Eq.(\ref{eq:FACT})
\bea
\Phi_0=\left({1   \over  \bk^{2}+\varepsilon^{2}}-
{1 \over  (\bk-\bkappa )^{2}+\varepsilon^{2}}\right)\nonumber\\
\bPhi_1=\left({\bk \over  \bk^{2}+\varepsilon^{2}} - 
{\bk-\bkappa  \over  (\bk-\bkappa )^{2}+\varepsilon^{2}}\right),
\label{eq:PHI01}
\eea
 and
\beq
\varepsilon^2=z(1-z)Q^{2}+(1-z)m_{t}^{2}+zm_b^2.
\label{eq:VAREPS}
\eeq
Then, making use of the technique
developed in \cite{IN2003} for electro-production processes,  
separate the $\bkappa^{2}$-integration in (\ref{eq:FACT}) into 
the soft, 
\beq
\bkappa^{2} \lsim \overline{k^2}\equiv\varepsilon^{2}+
{\bm k}^{2}
\label{eq:SOFT}
\eeq
 and hard, 
\beq
\bkappa^{2} \gsim \overline{k^2},
\label{eq:HARD}
\eeq
regions of the gluon momentum. For soft gluons, in the DGLAP \cite{DGLAP}
 region,  upon the 
azimuthal 
integration
we get 
\bea
\int {d\varphi\over 2\pi} \bPhi_1^2\approx
\bkappa ^2{ \varepsilon^{4} + (\bk^2)^2 \over (\bk^{2}+
\varepsilon^{2})^4}\nonumber\\
\int {d\varphi\over 2\pi} \Phi_0^2\approx
{2\bkappa^2\bk^2 \over (\bk^{2}+\varepsilon^{2})^4}
\label{eq:PHINT}
\eea

In Eq.(\ref{eq:FACT}) the QCD running coupling $\alpha_{S}(q^{2})$ 
enters 
the integrand at the largest
relevant virtuality,
$q^2={\rm max }\{\overline{k^2},\bkappa ^{2}\}.$
For soft gluons one can take
$
q^{2}=\overline{k^2}$
and  we arrive at the fully differential
distribution of the t-quark in $z$ and $\bk$,
 \bea
{d\sigma_{\lambda}(x,Q^{2})\over dz d^{2}\bk }\approx
\alpha_{W} \alpha_{S}(\overline{k^2})
G(x_g,\overline{k^2})\nonumber\\
\times\left[S_{\lambda}{2\bk^2 \over (\bk^{2}+\varepsilon^{2})^4}+
P_{\lambda}
{\varepsilon^{4} + 
(\bk^2)^2 \over (\bk^{2}+\varepsilon^{2})^4}\right], 
\label{eq:DSMAIN}
\eea
where $\overline{k^2}$ comes from Eq.(\ref{eq:SOFT}).
Integrating over $\bk$ yields 
\bea
{d\sigma_{\lambda}(x,Q^{2})\over dz}\approx
{2\pi\alpha_{W} \over 3 \varepsilon^2}
\left({S_{\lambda}\over 2\varepsilon^2}
+P_{\lambda}\right)\alpha_{S}(\varepsilon^2)G(x_g,\varepsilon^2).
\label{eq:DSDZ}
\eea 
The leading contribution to Eq.(\ref{eq:DSDZ}) comes  from 
$\bk^2\lsim \varepsilon^2/3$. We factored out the slowly varying product
$\alpha_{S}G$ at $\overline{k^2}\approx\varepsilon^2$.
The $\bk$- and $\bkappa$-dependence of $x_g$ is 
given by Eqs.(\ref{eq:XG},\ref{eq:M2}).  

Then, going from ${d\sigma_{L}/dz}$ to the 
 structure function of the nucleon probed by  the 
left-handed ($\lambda=L$) gauge boson, $F_L(x,Q^2)$ 
(see Eq.(\ref{eq:FTOSIG})),
we find  the soft gluon contribution to the P-wave 
component\footnote{Hereafter, the upper index in  $F_{\lambda}^{L}$ 
corresponds to the angular momentum
$L=0,1$ of the $q\bar q^{\prime}$-state which the 
light-cone $W$-boson transforms into.} 
of 
\beq
F_L=F_L^1+F_L^0
\label{eq:FL10}
\eeq
denoted by $F^1_{L}$
\bea
F^1_{L}(x,Q^{2}
\approx{2Q^2}\int_{0}^{1}
{dzz^2\over \varepsilon^{2}} 
 {\alpha_S(\varepsilon^{2}) \over 3\pi}G(x_g,\varepsilon^2).
\label{eq:FL1}
\eea
 The leading contribution to 
$F^1_L$ comes from 
\beq 
z\sim 1-{m_b^2\over m_t^2+Q^2}.
\label{eq:zlead}
\eeq 
Therefore,  Eq.(\ref{eq:FL1}) can be approximated by
\bea
F^1_{L}(x,Q^{2})\approx
{2Q^2\over {m_t^2+Q^2}}\int_{m_b^2}^{C(m_t^2+Q^2)}
{d\varepsilon^{2}\over \varepsilon^{2}} 
 {\alpha_S(\varepsilon^{2}) \over 3\pi}G(x_g,\varepsilon^2),
\label{eq:FLP}
\eea
where $C\approx 0.25$. Certainly,
$\sigma^{\nu N}$ is dominated by $Q^2\ll m_t^2$.
However, $F_L^1$ presented in the form (\ref{eq:FLP})
allows straightforward extension  to the processes 
induced by the charm-strange current, where $Q^2\gg m_c^2$ and $x_g\sim 2x$.

The  S-wave component of $F_L$, denoted by $F^0_L$,
is very small compared to $F_L^1$. Indeed,
 the $z$-integration in $F_L^0$
 converges rapidly at $z\sim 1-m_b^2/(m_t^2+Q^2)$
and
\bea
F_L^0(x,Q^2)\approx {Q^2m_b^2}\int_0^1{dz z^2\over \varepsilon^4}
{\alpha_S(\varepsilon^2)
\over 3\pi}
G(x_g,{\varepsilon^{2}})\nonumber\\
\approx {Q^2\over m_t^2+Q^2}
{\alpha_S(m_b^2)\over 3\pi}
G(x_g,m_b^2).
\label{eq:FLS}
\eea
Therefore, $F_L^0\ll F_L^1$ and $F_L\approx F_L^1$.

The $S$-wave component  of the  
right-handed structure function  
\beq
F_R=F_R^0+F_R^1
\label{eq:FR10}
\eeq
is as follows
\bea
F_R^0(x,Q^2)\approx {Q^2m_t^2}\int_0^1{dz (1-z)^2\over 
\varepsilon^4}
{\alpha_S(\varepsilon^2)
\over 3\pi}
G(x_g,{\varepsilon^{2}})\nonumber\\
\approx{Q^2\over m_t^2}
{\alpha_S(m_t^2/2)\over 3\pi}G(x_g,m_t^2/2).
\label{eq:FRS1}
\eea 
At $Q^2\ll m_t^2$ the ratio ${(1-z)^2/\varepsilon^4}$ in 
Eq.(\ref{eq:FRS1}) is
flat in $z$  and $\langle z \rangle \sim 1/2$.

For $Q^2\gg m_t^2$  the structure function  $F_R^0$ is dominated by  
$z\sim m_t^2/Q^2$ and $x_g\approx 2x$. Hence,
\beq
F_R^0(x,Q^2)\approx {\alpha_S(2m_t^2)\over 3\pi}G(x_g,2m_t^2).
\label{eq:FRS2}
\eeq

The P-wave component of  $F_R$  reads
\bea
F_R^1(x,Q^2)\approx {2Q^2}\int_0^1{dz (1-z)^2\over \varepsilon^2}
{\alpha_S(\varepsilon^2)
\over 3\pi}
G(x_g,{\varepsilon^{2}}).
\label{eq:FRP1}
\eea
For $Q^2\ll m_t^2$, $(1-z)^2/ \varepsilon^2\approx (1-z)/m_t^2$ and 
\bea
F_R^1(x,Q^2)\approx {Q^2\over m_t^2}
{\alpha_S(m_t^2)\over 3\pi}G(x_g,m_t^2).
\label{eq:FRP2}
\eea
\begin{figure}[h]
\psfig{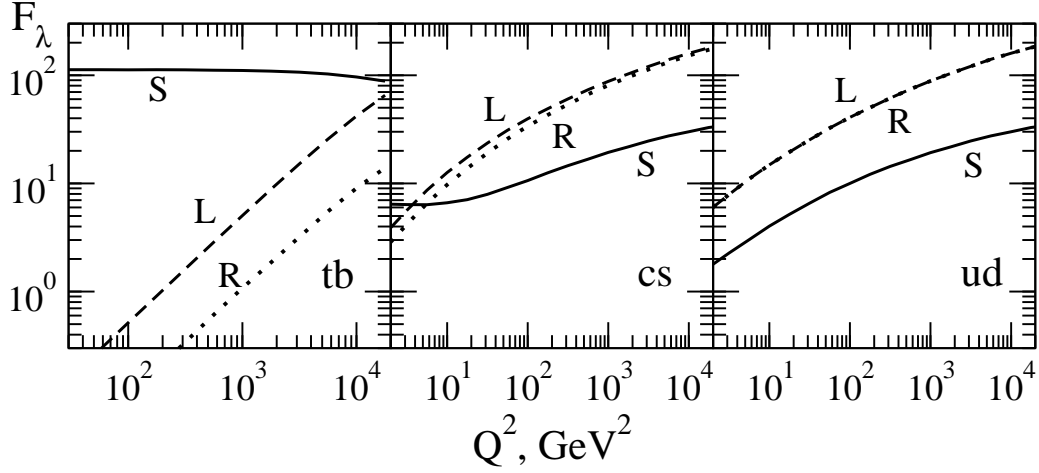}
\vspace{-0.5cm}
\caption{The nucleon structure function $F_{\lambda}$ for $\lambda=S,L,R$
as a function of $Q^2$ at $x=5.10^{-7}$ in  the neutrino reactions induced by
the  top-bottom, charm-strange and 
up-down quark currents.} 
\label{fig:fig1}
\end{figure}  
In the region of very high virtualities of the gauge boson, 
$CQ^2\gg m_t^2$ (see Eq.(\ref{eq:FLP})),
\bea
F_R^1(x,Q^2)\approx 2\int_{m_t^2}^{CQ^2}
{d\varepsilon^2\over \varepsilon^2}
{\alpha_S(\varepsilon^2)
\over 3\pi}
G(x_g,{\varepsilon^{2}})
\label{eq:FRP3}
\eea
with $x_g\approx 2x$.
Certainly,  Eq.(\ref{eq:FRP3}) is irrelevant to the problem of
$\sigma^{\nu N}$: the latter is dominated by $Q^2\ll m_t^2$.  However,  
Eq.(\ref{eq:FRP3}) is not entirely  useless. 
It describes - with obvious substitutions -
 the dominant contribution 
to the $F_R^1$ in high-$Q^2$ processes  induced by the charm-strange current 
(see Fig.\ref{fig:fig1}). 

From  explicite expressions for $F_L$ and $F_R$ obtained above 
it is 
evident that for the top-bottom current
\beq
F_L\gg F_R.
\label{eq:FLFR}
\eeq 
in a wide range of $Q^2$.
Another observation is that
the contribution of left-handed  $W$-bosons
to $\sigma^{\nu N}$ in processes induced by the top-bottom current
is much smaller than that  coming from the absorption of scalar  W-bosons
\beq
F_S\gg F_L.
\label{eq:FLFS}
\eeq
The scalar structure function $F_S$ (under the name longitudinal) has 
been discussed in \cite{UHE10}.   It was found that
the higher twist corrections
brought about by  the non-conservation  of the  top-bottom current 
result in considerable enhancement of the $P$-wave component of 
\beq
   F_S=F_S^1+F_S^0,
\label{eq:FSUM}
\eeq
denoted by $F^1_{S}$
\bea
F^1_{S}(x,Q^{2})\approx
{m_t^2\over {m_t^2+Q^2}}\int_{m_b^2}^{\varepsilon_m^2}
{d\varepsilon^{2}\over \varepsilon^{2}} 
 {\alpha_S(\varepsilon^{2}) \over 3\pi}G(x_g,\varepsilon^2),
\label{eq:FSP}
\eea
where $\varepsilon_m^2=m_t^2$ if $Q^2<m_t^2-m_b^2$ and 
$\varepsilon_m^2\approx (Q^2+m_t^2)/4$ if $Q^2 > m_t^2-m_b^2$.

  At $Q^2\ll m_t^2$,   $F_S^0$ is much smaller than  $F_S^1$. 
Indeed, from Eq.(\ref{eq:DSDZ})
it follows that
\bea
F_S^0\approx {Q^2\over 12\pi}\int_0^1 {dzS_S\over \varepsilon^4}
\alpha_S(\varepsilon^2)G(x_g,\varepsilon^2),
\label{eq:FSS0}
\eea
where $S_S$ comes from Eq.(\ref{eq:SPS}) and $\varepsilon^2$ from 
Eq.(\ref{eq:VAREPS}). 
For $Q^2\ll m_t^2$
\beq  
{Q^2S_S\over \varepsilon^4}\approx 2\left(1+{\delta^2\over(1-z+\delta^2)^2}\right)
\label{eq:LOWQ2}
\eeq
with $\delta=m_b/m_t$.
Therefore, 
\bea
F_S^0(x,Q^2)\approx {\alpha_S(m_t^2/2)
\over 6\pi}
G(x_g,m_t^2/2)\nonumber\\
+{\alpha_S(m_b^2)
\over 6\pi}
G(x_g,m_b^2).
\label{eq:FSS1}
\eea
Our $F_S$ survives the limit $m_t^2\to \infty$
{\footnote{The limit $m_t^2\to \infty$ implies the limit of the infinite
 neutrino energy, $1/x_g\to \infty$. 
The finite energy effects, built in the gluon density 
function , suppress the structure function in Eq. (\ref{eq:FSS1}).
Specifically, for  $x_g\to 1$ the gluon density $G$ is known to vanish 
as $(1-x_g)^n$, where $n\approx 5$ and $x_g$ comes from 
Eqs.(\ref{eq:XG}, \ref{eq:M2}).}}.
The point is that the scalar/longitudinal W-boson 
interacts with  the quark  current $j_{\mu}=V_{\mu}-A_{\mu}$ 
which  is not conserved. The  vertex  $W\to t{\bar b}$ which  
  is  $\propto \partial_{\mu}j_{\mu}\propto m_t\pm m_b$ gives the factor $m_t^2$
in the expression for $F_S$ and cancels $1/m_t^2$ coming from the $tb$ quark box.
Evidently, there is no real clash between
 Eq.(\ref{eq:FSS1}) and  the decoupling theorem \cite{AC}. 
The latter is relevant to  calculating 
observables, for instance cross sections in 
kinematical regions well below certain heavy quark thresholds.  We are 
dealing with the real top production by neutrinos of ultra-high energy,
though at  $Q^2\ll m_t^2$. 
\begin{figure}[h]
\psfig{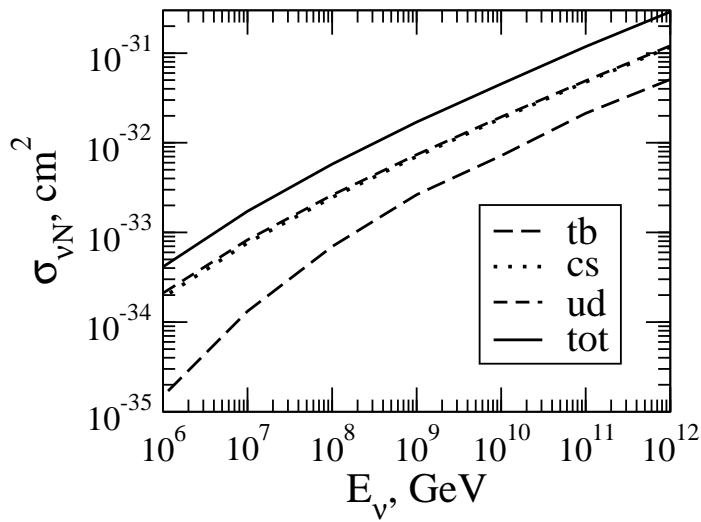}
\vspace{-0.5cm}
\caption{The total $\nu N$ charged current cross section (solid line)
 decomposed into 
components of different origin as a function of the 
laboratory neutrino energy. } 
\label{fig:fig2}
\end{figure}  

Two terms in Eq.(\ref{eq:FSS1}) correspond to two very different kinematics.
The first one describes  the symmetric $t\bar b$-state with uniform
 $z$-distribution  and ${\langle z\rangle}\approx 1/2$,
the second one corresponds to the asymmetric configuration with 
 $z\approx 1-\delta^2$. Both $t\bar b$-states have 
approximately the same invariant mass, $M^2\sim 2m_t^2$, and, 
consequently,
probe the gluon density at nearly the same $x_g$.

For  $Q^2\gg m_t^2$ -- the substitution $t\to c$ suggests itself --
the ratio $S_S/\varepsilon^4$ is flat in $z$ and 
  $Q^2S_S/\varepsilon^4\approx 8$ with ${\langle z\rangle}\approx 1/2$ and $x_g\approx 2x$. 
 Consequently, 
\bea
F_S^0(x,Q^2)\approx {2\alpha_S({Q^2/4})
\over 3\pi}
G(x_g,Q^2/4).
\label{eq:FSS2}
\eea
 In the neutrino
reactions induced by the charm-strange current the term $F_S^1$ is suppressed
at high $Q^2\gg m_c^2$ by the factor $\sim m_c^2/Q^2$ and  the 
scalar/longitudinal  structure function $F_S$ is 
dominated by its S-wave component,  $F_S^0$ 
\cite{FZCS1,FZCS2}. Specific features of the structure functions 
$F_{\lambda}$  are illustrated by  Fig.\ref{fig:fig1}. From 
Fig.\ref{fig:fig1} and Eqs. (\ref{eq:DSDXDY},\ref{eq:Q2W2}) it follows
that the left-right asymmetry generated by the top-bottom current is 
much stronger than that generated by the charm-strange current  but
the top-bottom contribution to the 
total neutrino nucleon cross section is dominated by the absorption of the
 scalar gauge boson.

 In our numerical
studies of $F_{\lambda}$ and $\sigma^{\nu N}$ we rely upon the color dipole 
factorization represented by the first line of Eq.(\ref{eq:FACT}), where
the total cross section of interaction of the $t\bar b$ color dipole
of the transverse size ${\bf r}$ with the nucleon target is related to 
the differential density of gluons ${\cal F}(x_g,\kappa^2)$ by the 
equation \cite{NZDLLA}
\bea
\sigma(x,r)={\pi r^2\over N_c}\int {d^2\bkappa\over \kappa^2}
{4[1-\exp(i\bkappa\br)]\over \kappa^2r^2}\alpha_S(\kappa^2)
{\cal F}(x_g,\kappa^2).
\label{eq:SIGMA}
\eea
 The light cone  density of the $t\bar b$ Fock states with the 
transverse 
size ${\bf r}$ and the fraction $z$  of the
$W$-boson
light cone momentum carried by the $t$-quark is \cite{FZ1,Kolya92}
\beq
|\Psi_{\lambda}(z,r)|^2={2\alpha_WN_c\over (2\pi)^2}
\left[S_{\lambda}K_0^2(\varepsilon r)+
P_{\lambda}\varepsilon^2K_1^2(\varepsilon r)\right],
\label{eq:PsiLR}
\eeq
where $S_{\lambda}$ and $P_{\lambda}$ come from
 Eqs.(\ref{eq:SPLR},\ref{eq:SPS}).
In (\ref{eq:PsiLR}) $K_{0,1}(y)$ is the modified Bessel 
function.
The terms proportional to $K_0^2(\varepsilon r)$ and $K_1^2(\varepsilon r)$ 
describe  the 
quark-antiquark states with the angular momentum $L=0$ (S-wave) and $L=1$
(P-wave), respectively. The $\log(1/x)$-evolution of $\sigma(x,r)$ is
determined by  the infrared regulated  BFKL equation with running
 coupling \cite{NZZJL94,NZJETP94}. 
The
preferred choice of the infrared regularization  gives
the  intercept of the pomeron trajectory, $\alpha_{\Pom}(t)$,
 in the angular momentum 
plane $\Delta_{\Pom}=\alpha_{\Pom}(0)-1=0.4$  and leads to a very good
description  of the  data on the proton 
structure functions
at small $x$ \cite{CharmBeauty}. 
In particular, this means 
that the gluon density and  all structure functions  grow 
fast with growing $1/x$, $F_{\lambda}\propto x^{-\Delta_{\Pom}}$.
The analysis of the unitarity effects will be published elsewhere.
The structure functions of the  isoscalar nucleon probed by $W$-bosons
of different helicity  
are presented in Fig.\ref{fig:fig1}.
 
 The charged current 
 neutrino-nucleon cross section, $\sigma^{\nu N}$ as a 
function of the neutrino energy, 
$E_{\nu}$ and its decomposition into components of different origin 
is shown in  Fig.\ref{fig:fig2}.
 
Summarizing,  we presented
  the $\bkappa$-factorization formulas  for the  differential
 cross section 
$d\sigma_{\lambda}/dz d^{2}\bk$ which describes  the absorption of
scalar, left-handed and right-handed  W-bosons.  We isolated
leading contributions to the related structure functions
 $F_{\lambda}(x,Q^2)$ at low $x$ and 
high $Q^2$ and  obtained  simple and numerically accurate estimates
 for  $F_{\lambda}$ in processes induced by  massive 
quark currents. It was shown that the 
non-conservation  of the  top-bottom current leads to the hierarchy of 
 structure functions $F_S\gg F_L\gg F_R$. Making use of 
 the color dipole BFKL approach to the  $\log(1/x)$ QCD evolution we
evaluated  the charged current $\nu N$ cross sections 
for ultra-high energy neutrino beams.

{\bf Acknowledgments.} 
V.R.~Z. thanks 
 the Dipartimento di Fisica dell'Universit\`a
della Calabria and the Istituto Nazionale di Fisica
Nucleare - gruppo collegato di Cosenza for their warm
hospitality while a part of this work was done.
The work was supported in part by the Ministero Italiano
dell'Istruzione, dell'Universit\`a e della Ricerca and  by
 the RFBR grants 09-02-00732 and 11-02-00441.

\end{document}